\documentclass[a4paper,12pt]{article}

\usepackage{amsthm,amssymb,amsmath,url,amsfonts,mathrsfs}

\theoremstyle{plain}

\usepackage{graphicx}
\usepackage{dcolumn}
\usepackage{bm}
\newcommand{\ra}{\rangle}
\newcommand{\la}{\langle}
\newcommand{\R}{\mathbb{R}}
\newcommand{\X}{\mathbb{X}}

\newcommand{\C}{\mathbb{C}}

\newcommand{\XX}{{\cal X}}
\renewcommand{\SS}{{\cal S}}
\newcommand{\EE}{{\cal E}}
\renewcommand{\AA}{{\cal A}}

\newcommand{\MM}{{\cal M}}
\renewcommand{\SS}{{\cal S}}
\newcommand{\HH}{{\cal H}}
\newcommand{\YY}{{\cal Y}}

\newcommand{\DD}{{\cal D}}

\newcommand{\x}{{\bf x}}

\begin{document}
\title{\bf On the preferred-basis problem and its possible solutions}
\author{Bruno Galvan \footnote{e-mail: b.galvan@virgilio.it}\\ \small Via di Melta 16, 38121 Trento, Italy.}


\date{\today}
\maketitle
\begin{abstract}
The preferred basis problem is mentioned in the literature in connection with the measurement problem and with the Many World Interpretation. It is argued that this problem actually corresponds to two inequivalent problems: (i) the preferred-decomposition problem, i.e., what singles out a preferred {\it decomposition} of a suitable state vector of a system as the sum of a finite or countable set of vectors?, and (ii) the preferred-representation problem, i.e., what singles out a preferred {\it representation} for the Hilbert space of a system? In this paper the preferred-decomposition problem is addressed and two processes, namely decoherence and permanent spatial decomposition (PSD), are examined and compared as possible solutions to this problem. It is shown that, perhaps contrary to common belief, in realistic situations decoherence is not sufficient to solve the preferred-decomposition problem. PSD is the (hypothesized) tendency of the wave function of the universe to decompose into permanently non-overlapping wave packets. Three phases can be roughly identified as composing PSD: Microscopic decomposition, amplification, and interaction with the environment. Decoherence theory considers only the interaction with the environment and ignores the first two phases. For this reason PSD is fundamentally different from decoherence and, unlike decoherence, provides a simple and non-elusive solution to the preferred-decomposition problem.
\end{abstract}


\section{The preferred-basis problem} \label{preferred}

The preferred basis problem is mentioned in the literature in connection with the measurement problem and with the Many World Interpretation (MWI). In the first case it refers to the problem of how to decompose the post-measurement state of microscopic system + apparatus into the sum of states corresponding to the various possible outcomes of the measurement \cite{schloss2}, while in the second case it refers to the problem of how to decompose the wave function of the universe into the sum of states corresponding to the different worlds \cite{mwistanford}. This is better explained in the section \ref{does}. There is however a certain ambiguity in regard to exactly what the preferred-basis problem is. Consider the following examples.

In Section 2.5.2 of the book \cite{schloss2}, the preferred-basis problem in the measurement process is presented as (a) the non-uniqueness of the Schmidt decomposition of the post-measurement state of microscopic system + apparatus. On the other hand, on page 50 of the same book the preferred-basis problem is defined as follows: (b) ``What singles out the preferred physical quantities in nature--e.g., why are physical systems usually observed to be in definite positions rather than in superpositions of positions?''. Definitions (a) and (b) are clearly not equivalent, at least from a formal point of view.

Another example: in \cite{everettbarrett} one can read that (c) ``there are always many ways one might write the quantum-mechanical state of a system as the sum of vectors in the Hilbert space; in choosing a preferred basis, one chooses a single set of vectors that can be used to represent a state and thus one chooses a single {\it preferred} way of representing a state as the sum of vectors in the Hilbert space''. On the other hand, in the same text one can also read: (d) ``The problem of choosing which observable to make determinate is known as the preferred-basis problem''.

The texts quoted suggest that there are arguably two inequivalent versions of the preferred-basis problem. In order to obtain a more precise formulation, let us recall the two notions of {\it decomposition} and of {\it representation}. Let $\HH_\SS$ be the Hilbert space of a system $\SS$. In the context of the measurement problem $\SS$ is the compound system consisting of the microscopic system + apparatus, while in the context of the MWI it is the whole universe. A {\it decomposition} of a vector $|S \ra \in \HH_\SS$ can be defined as a finite or countable set of linearly independent vectors $\{|S_1 \ra, |S_2 \ra, \ldots \}$ such that $\sum_i |S_i \ra= |S \ra$ (see section \ref{formal}). On the other hand a {\it representation} for $\HH_\SS$ is an orthonormal or generalized basis of $\HH_\SS$. A representation is typically defined by a complete set of commuting observables, whose proper or generalized eigenvectors constitute the representation. For example, position operators are a complete set of commuting observables for a system of $N$ spinless particles, and they define the generalized basis $\{|x \ra\}_{x \in \R^{3N}}$. The two notions of decomposition and of representation are obviously inequivalent. In fact, an orthonormal basis also defines a decomposition for any vector of $\HH_\SS$, but this is no longer true for a generalized basis as, for example, position representation. On the other hand, a generic decomposition of a given vector certainly does not form a basis. Note that a decomposition refers to a specific vector, while a representation refers to the whole Hilbert space. Since a vector usually evolves with time, a decomposition may evolve with time as well, while a representation is typically fixed.

Therefore, according to the texts quoted above, it appears that at least the following two versions of the preferred-basis problem can be formulated:
\begin{itemize}
\item[A.] {\it The preferred-decomposition problem}: what singles out a preferred decomposition of a suitable vector $|S \ra \in \HH_\SS$?
\item[B.] {\it The preferred-representation problem}: what singles out a preferred representation for $\HH_\SS$?
\end{itemize}
The quoted definitions (a) and (c) arguably correspond to the preferred-decomposition problem, while the quoted definitions (b) and (d) appear to correspond to the preferred-representation problem. In fact, singling out a set of physical quantities means singling out a (complete) set of commuting observables.

In this paper only the preferred-decomposition problem will be addressed because, as explained in section \ref{does}, it is the most directly connected with the measurement problem and with the problem of defining the worlds in the MWI. In particular, two possible solutions to the preferred-decomposition problem will be discussed, namely (i) decoherence and (ii) permanent spatial decomposition (PSD). Decoherence is probably the mainstream solution to the preferred-decomposition problem \cite{Zurek2003, schloss2, mwistanford, Tegmark2010, Wallace2003}, even though the limits of this solution have been emphasized \cite{Zeh1993, Stapp2002}. In this paper the usual reasoning leading to the conclusion that decoherence solves the preferred-decomposition problem is reviewed, and its weakness is emphasized. PSD is the (hypothesized) structural tendency of the wave function of the universe to decompose into the sum of permanently non-overlapping parts. The name PSD has been proposed for the first time by the author \cite{Galvan2009}, but this process has already been considered and studied in the literature, mainly in connection with the de Broglie-Bohm theory \cite{Bohm:1951, Bohm:1951xx, Bohm:1987np, undivuni, qequilibrium, Struyve:2006, Peruzzi:1996jy} but not only \cite{galvan:4, zanghimw}. PSD is therefore mainly known and considered in the rather restricted community of Bohmian physicists, and for this reason it is not often explicitly mentioned as a possible solution of the preferred-basis problem. As in decoherence theory, the interaction with the environment (IWE) plays an important role in the process of PSD, but it will be shown that this process also includes two dynamical phases which are not considered in decoherence theory, namely microscopic decomposition and amplification. For this reason PSD is fundamentally different from decoherence, even though these two phenomena are not always clearly differentiated in the literature. This will be discussed in section \ref{process}, where it will be shown that, unlike decoherence, PSD provides a non-elusive solution to the preferred-decomposition problem.

\section{Decoherence} \label{decoherence}

Decoherence theory \cite{Breuer2002, Giulini1996, schloss2} studies the evolution of a system $\SS$ interacting with an environment $\EE$. The evolution of $\SS$ is represented by the reduced density matrix
\begin{equation}
\rho_\SS(t):=\hbox{Tr}_\EE\{U(t)|\Psi_0 \ra \la \Psi_0 | U^\dagger(t)\},
\end{equation}
where $|\Psi_0 \ra$ and $U(t)$ are the initial state and the time evolution operator of $\SS + \EE$, respectively. Typically the system is mesoscopic or macroscopic, and the environment has a very large number of degrees of freedom, whose detailed knowledge is not interesting for the experimenter. Decoherence theory studies in particular those interactions in which the environment has a small dynamical influence on the system; for example, if the system is a mesoscopic particle which scatters environmental particles, decoherence theory is mainly concerned with the situations in which the environmental particles do not significantly alter the momentum of the mesoscopic particle. In these situations the main effect of the environment on the system derives from the pure quantum mechanical phenomenon of entanglement.

In this section the most relevant physical and mathematical features of decoherence theory will be exemplified by considering three different theoretical models of decoherence, namely (i) the ideal decoherence model, (ii) the damped harmonic oscillator and (iii) the collisional decoherence model. 

\vspace{3mm}
{\bf The ideal decoherence model} (\cite{ Breuer2002}, section 4.1). This model represents a highly idealized situation, and it is useful mainly because it well exemplifies the main conceptual features of decoherence. This model is defined by the following three dynamical rules:
\begin{itemize}
\item[1.] {\it Pointer states}: There exists a set $\{|S_i \ra \}_{i \in I}$ of states of $\SS$ which are stable under time evolution, that is 
\begin{equation} \label{stab}
U(t) |S_i \rangle |E \rangle = |S_i \rangle |E_i (t) \rangle
\end{equation}
for $t \geq 0$ and for any state $|E \rangle $ of $\EE$. The states $\{|S_i \ra \}_{i \in I}$ are referred to as {\it pointer states}.
\item[2.] {\it Pointer basis}: The pointer states form an orthonormal basis $\{|S_1 \ra, |S_2 \ra, \ldots\}$ of the Hilbert space of $\SS$; this basis is referred to as the {\it pointer basis}.
\item[3.] {\it Decoherence:} The states $\{|E_1(t) \rangle, |E_2(t) \ra, \ldots \}$ have the property that $\la E_i(t) |E_j (t) \ra \approx \delta_{i j}$ for $t \gg \tau_D$, where $\tau_D$ is a suitable time scale which is referred to as the {\it decoherence time scale}.
\end{itemize}
The name {\it decoherence} for property 3 has been chosen for the following reason. If $|S \ra |E \ra = \left (\sum_i c_i |S_i \ra \right ) |E \ra$ is the initial state of $\SS + \EE$, then
\begin{equation}
 \rho_\SS(0) = |S \ra \la S| = \sum_{i, j}c_i c_j^* |S_i \ra \la S_j| \to \rho_\SS(t) \approx \sum_i |c_i|^2 |S_i \ra \la S_i | \;  \hbox{ for } \; t \gg \tau_D.
\end{equation}
In words, the initial pure state density matrix $\rho_\SS(0)$ evolves to the mixed density matrix $\rho_\SS(t)$, which is diagonal in the pointer basis and time-independent. The coherence between different pointer states is therefore suppressed by IWE.

Rules 1 and 3, namely the existence of stable pointer states and the suppression of coherence between orthogonal pointer states are the most relevant effects of IWE studied in decoherence theory, and they are present in practically all the models of decoherence. However in more realistic models there is no state exactly satisfying the condition (\ref{stab}), and therefore in such models the pointer states are defined as the states which satisfy ``as much as possible'' that condition. The exact mathematical form in which the requirement ``as much as possible'' is expressed, and therefore the exact mathematical definition of pointer states, depends on the specific decoherence model. In what follows the name ``pointer state'' will be reserved for the states for which a precise mathematical definition has been given, while the states which satisfy approximately the condition (\ref{stab}), namely that evolve slowly compared to the decoherence time scale and become few entangled with the environment, will be generically referred to as {\it robust} states.

Contrary to rules 1 and 3, rule 2 has no general validity, not even in an approximate form. In other words, in more realistic models of decoherence the pointer states typically form an overcomplete basis rather than an orthonormal basis. This is well exemplified by the second decoherence model we will examine.

\vspace{3mm}
{\bf The damped harmonic oscillator}. In order to study this model, it is necessary to introduce the coherent states of the harmonic oscillator \cite{Glauber1963}. Let $H_\SS= \omega a^\dagger a$ be the usual Hamiltonian of a harmonic oscillator, where $a^\dagger$ and $a$ are the creation and annihilation operators, respectively. The coherent states $\{|\alpha \rangle\}_{\alpha \in \C}$ are the normalized eigenvectors of the non-Hermitian operator $a$, that is $a |\alpha \ra = \alpha |\alpha \ra$. They are not orthogonal: $\la \alpha | \beta \ra= \exp(-|\alpha|^2 /2 - |\beta|^2/2 + \alpha \beta^*)$, and they form an overcomplete set of states satisfying the completeness relation $\frac{1}{\pi} \int |\alpha \ra \la \alpha | d^2 \alpha = I$.

Let us now consider the damped harmonic oscillator at zero temperature (\cite{ Breuer2002}, Section 4.4). In this case the reduced density matrix satisfies the following master equation:
\begin{equation} \label{mast}
\dot \rho_\SS(t) = \left (- i \omega - \frac{\gamma}{2} \right ) a^\dagger a \rho_\SS(t) + 
\left (i \omega - \frac{\gamma}{2} \right ) \rho_\SS(t) a^\dagger a +
\gamma a \rho_\SS(t) a^\dagger.
\end{equation}
The solution of (\ref{mast}) with initial value $\rho_\SS(0)=|\alpha \ra \la \beta|$ is 
\begin{equation} \label{sol}
\rho_\SS(t)=f(t) |\alpha_t \ra \la \beta_t|,
\end{equation}
where
\begin{eqnarray}
& & \alpha_t= \alpha \exp (-i \omega -  \gamma t /2), \nonumber \\
& & \beta_t= \beta \exp (-i \omega -  \gamma t /2), \\
& & f(t)= \la \beta | \alpha \ra^{[1 - \exp(- \gamma t)]}. \nonumber
\end{eqnarray}
From (\ref{sol}) one deduces that if $\rho_\SS(0)=|\alpha \ra \la \alpha |$ then $\rho_\SS(t)=|\alpha_t \ra \la \alpha_t|$, that is coherent states remain pure under time evolution; for this reason they are identified as the pointer states of the damped harmonic oscillator. Moreover, if $\rho_\SS(0)=|S \ra \la S|$, with $|S \ra= c_1 |\alpha \ra + c_2 |\beta \ra$, then
\begin{equation}
\rho_\SS(t)= |c_1|^2 |\alpha_t \ra \la \alpha_t | + |c_2|^2 |\beta_t \ra \la \beta_t | + c_1 c_2^* f(t) |\alpha_t \ra \la \beta_t | + c_1^* c_2 f^*(t)|\beta_t \ra \la \alpha_t |.
\end{equation}
For $ t \gg \gamma^{-1}$ we have $f(t) \approx \la \beta | \alpha \ra$, and therefore if $|\la \alpha | \beta \ra| \approx 0$ the interference terms of $\rho_\SS(t)$ are suppressed in the long time limit.

By comparing the properties of the damped harmonic oscillator with the three dynamical rules of the ideal decoherence model we can reach the following conclusions: (1) robust states still exist, and the pointer states can be reasonably identified with the coherent states; (2) the pointer states do not form an orthonormal basis; (3) pointer states diagonalize the reduced density matrix of the system (i.e., the coherence among pointer states is suppressed by IWE) only if the initial state of the system is the superposition of (quasi) orthogonal coherent states.

Let me remark in particular in regard to the following two consequences of the fact that the pointer states do not form an orthonormal basis: (i) they do not define any decomposition for a generic vector of the system and, (ii) contrary to what is perhaps commonly considered as a general property of pointer states, they do not diagonalize a generic density matrix of the system\footnote{Unfortunately this subject is often referred to with a certain ambiguity or approximation in the literature. For example, in \cite{Zurek1998} one can read: ``Einselection chooses a preferred basis in the Hilbert space in recognition of its predictability. That basis will be determined by the dynamics of the open system in the presence of environmental monitoring. It will often turn out that it is overcomplete. Its states may not be orthogonal, and, hence, they would never follow from the diagonalization of the density matrix''. Nevertheless, in a recent paper of the same author \cite{Zurek2010}, one can read: ``In this sense, pointer states are distinguished not only for forming the stable basis in which the density matrix of the system diagonalizes but also for being redundantly copied into the environment''.}.

\vspace{3mm}
{\bf The model of collisional decoherence}. In this model the environment is composed of a large number of environmental particles (air molecules or photons) which are scattered by the system \cite{Joos1985, Giulini1996, schloss2}. Here I will not enter into the mathematical details of this model, but will limit myself to saying that this model has features very similar to those of the damped harmonic oscillator. In this model the robust states of the system are spatially narrow wave packets, which of course are not necessarily orthogonal. The coherence between two wave packets is rapidly suppressed provided that the wave packets are spatially well separated. Recently, an exact definition of pointer states has been proposed for this model \cite{Busse2010}, and it has been shown that they form an overcomplete basis.

\section{Does decoherence solve the preferred-basis problem?} \label{does}

Since the famous 1981 paper of Zurek: {\it Pointer basis of quantum apparatus: Into what mixture does the wave packet collapse?} \cite{Zurek:1981}, the preferred-basis problem in the measurement process and its solution by means of IWE and decoherence is the introductory subject of many texts about decoherence (for example \cite{Zurek2003,schloss2}). Let us briefly review this.

In the situation of the measurement process the system $\SS$ is the compound system $\MM + \AA$, where $\MM$ is the microscopic system to be measured and $\AA$ the apparatus. Let us assume that an observable is measured possessing a basis of non-degenerate eigenvectors $\{|\phi_1 \ra, |\phi_2 \ra, \ldots \}$, and that $\MM$ is initially in the state $|\phi_1 \ra + |\phi_2 \ra$. According to the usual ideal measurement scheme, at the beginning of the measurement the state of $\SS$ is $(|\phi_1 \ra + |\phi_2 \ra ) |A_0 \ra$, where $|A_0 \ra$ is the ``ready'' state of the apparatus, and at the end of the measurement its state is
\begin{equation} \label{dec1}
|S \ra = |\phi_1 \ra |A_1 \ra + |\phi_2 \ra |A_2 \ra =:|S_1 \ra + |S_2 \ra,
\end{equation}
where $|A_i \ra$ is the state of the apparatus which has measured the outcome $i=1,2$. The preferred basis problem is presented as the fact that, other than with decomposition (\ref{dec1}), many other Schmidt decompositions exist for $|S \ra$, for example
\begin{equation}
|S \ra = \frac{1}{2}(|\phi_1\ra + |\phi_2 \ra)(|A_1 \ra + |A_2 \ra) + \frac{1}{2} (|\phi_1\ra - |\phi_2\ra)(|A_1 \ra - |A_2 \ra) =:|S'_1 \ra + |S'_2 \ra.
\end{equation}
The claim is usually made that, due to this ambiguity, the actual observable which is measured in the experiment is not defined until $\SS$ interacts with the environment. This interaction removes the ambiguity in favor of decomposition (\ref{dec1}), because the states $|S_1\ra$ and $ |S_2 \ra$ (but not the states $|S'_1 \ra$ and $|S'_2 \ra$) are pointer states and decohere, that is:
\begin{equation} \label{postm}
|S \ra |E \ra \to U(t) |S \ra |E \ra \approx |S_1 \ra |E_1(t) \ra + |S_2 \ra |E_2(t) \ra, \hbox{ with } \la E_i(t) | E_j(t) \ra \approx \delta_{ij} \hbox{ for } t \gg \tau_D.
\end{equation}
Equations (\ref{postm}) guarantee that the density matrix of $\SS$ is diagonalized by the states $|S_1\ra$ and $|S_2 \ra$. The conclusion is therefore reached that IWE and its effects, namely the definition of pointer states and decoherence, solves the preferred-basis problem in the measurement process.

At this point, three remarks are relevant. First of all we see that in this example the preferred-basis problem corresponds to the preferred-decomposition problem, because we are only interested here in finding a preferred decomposition of the specific vector $|S \ra$ and certainly not in finding a complete representation for the whole Hilbert space of $\SS$.

Secondly, as already noted by other authors \cite{Barvinsky1995, Rubin2004}, in this ideal model of measurement the decomposition $\{|S_1\ra, |S_2 \ra\}$ can be singled out without invoking IWE. In fact we can simply note that the states $|\phi_1\ra$ and $|\phi_2\ra$ (but not the states $|\phi_1\ra + |\phi_2\ra$ and $|\phi_1\ra - |\phi_2\ra$) of the microscopic system are stable with respect to the interaction with the apparatus, namely
\begin{equation}
|\phi_i \ra |A_0 \ra \to |\phi_i \ra |A_i \ra \hbox{ for } i=1, 2.
\end{equation}
This criterion of stability singles out the decomposition $\{|S_1 \ra, |S_2 \ra\}$, even though in this case stability is in relation to the interaction $\MM$-$\AA$ rather than to the interaction $\SS$-$\EE$. This criterion is however no longer applicable to less simple but more realistic models of the measurement process (see for example \cite{Ballentine1998}), and therefore the preferred-decomposition problem remains a real problem of the measurement process.

The third remark, which is the most important for us, is that IWE plays only a partial role in the solution of the preferred-decomposition problem, for the following reason. If the ideal decoherence model were applicable to the measurement process, IWE would define an orthonormal pointer basis for the system, and a preferred decomposition would be defined for a generic state $|S \ra$ of $\SS$. However, it is rather evident that the ideal model is not appropriate for a real measurement performed with real macroscopic apparatuses: For example, what would be the orthonormal pointer basis defined by IWE in the Hilbert space of, let us say, a photomultiplier? Note that such a space is the Hilbert space of the particles composing the photomultiplier, and the overwhelming majority of its states do not even remotely resemble a photomultiplier. The collisional model is certainly more appropriate for the measurement process; recall that in this model the pointer basis is an overcomplete (and therefore not orthonormal) basis, and the pointer or robust states are narrow wave packets. Thus IWE cannot single out a preferred decomposition of a generic state $|S \ra$ of $\SS$, and the decomposition $\{|S_1 \ra, |S_2 \ra\}$ can be singled out as preferred only if $|S_1 \ra$ and $|S_2 \ra$ are two narrow and spatially separated wave packets\footnote{This also corresponds to our usual intuitive image of the post-measurement state, where the states $|S_1 \ra$ and $|S_2 \ra$ represent the apparatus with two different macroscopic positions of the pointer.}. It is therefore necessary to hypothesize the existence of another process, preceding and independent of IWE, which ``prepares'' the state $|S \ra$ in the required superposition. It is evident that such a process, together with IWE, also has to be considered responsible for the existence of a preferred decomposition of $|S \ra$. This new process (actually two processes, as we will see) is a part of the general process of PSD that will be examined in the next section. The conclusion is therefore that the IWE alone cannot solve the preferred-decomposition problem in the context of the measurement process.

It may be useful to reformulate the above reasoning in a more general way and relate it to the experiments in which decoherence has been empirically tested. The previous reasoning allows the following conclusion: {\it in most realistic situations, including realistic measurement processes, the IWE does not define an orthonormal pointer basis for the system; as a consequence, IWE destroys coherence among robust states of the system only if the system has been previously prepared in a superposition of orthogonal or spatially separated robust states, and IWE plays no role in such a preparation phase}. For example, Zeilinger and coworkers have realized the controlled decoherence of a beam of $C_{70}$ molecules in a double-slit like experiment; in this experiment the initial superposition of spatially separated wave packets has been prepared by making the molecules pass through three free-standing gold gratings \cite{Zei:2004}. In another experiment Brune and coworkers have realized the controlled decoherence of a mesoscopic superposition of two quasi-orthogonal coherent states of the electromagnetic field in a cavity; in this case as well, the initial superposition has been prepared by means of a specific and clever experimental setup \cite{Brune1996}.

\vspace{3mm}
Let us now consider the preferred basis-problem in the context of MWI. In \cite{mwistanford} it is assumed that the quantum state of the universe is decomposable into the sum of worlds:
\begin{equation} \label{mw}
|\Psi_{UNIVERSE} \ra = \sum \alpha_i |\Psi_{WORLDi} \ra.
\end{equation}
The preferred-basis problem is presented as the problem of how to define the decomposition (\ref{mw}), and it is obvious that in this case this problem also corresponds to the preferred-decomposition problem. If we attempt to make use of pointer states for defining the decomposition (\ref{mw}) we encounter the same problems as in the case of the measurement process, with the following complication: The definition of pointer states is based on the decomposition of the global system into system + environment, and it is not at all clear how to perform such a decomposition when the global system is the universe; this is considered to be a severe conceptual difficulty of decoherence theory \cite{ Zurek1998, Schlosshauer:2003zy}. The conclusion is therefore that decoherence does not even solve the preferred-decomposition problem in the context of the MWI.

Nevertheless, decoherence also appears to be the mainstream solution in this situation \cite{mwistanford, Tegmark2010, Wallace2003}. There are however also some criticisms. For example, Zeh claims that (in the context of decoherence theory) ``The precise definition of the dynamically independent components (`branches') remains elusive'' \cite{Zeh1993}; also Stapp, on the basis of reasoning analogous to that presented here, concludes that decoherence does not solve the preferred-decomposition problem in MWI \cite{Stapp2002}.

\section{The process of permanent spatial decomposition} \label{process}

\begin{figure}
\begin{center}
\includegraphics {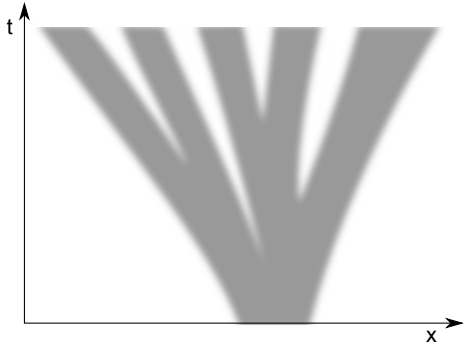}
\caption{Schematic representation of the process of permanent spatial decomposition. The gray zone is the support of the wave function; it is slightly fuzzy to emphasize the fact that the support can be only vaguely defined.} \label{fig1}
\end{center}
\end{figure}

The process of permanent spatial decomposition (PSD) is the (hypothesized) structural tendency of the wave function of a macroscopic system, typically the universe, to decompose into permanently non-overlapping parts, where ``non-overlapping'' is understood to mean ``approximately non-overlapping in configuration space''. Following Bohm \cite{undivuni}, the non-overlapping elements of the decomposition will be referred to as {\it channels}. Recall that for two state vectors representing the universe to be non-overlapping, it is sufficient that they differ with respect to the position of a single particle. After decomposition every channel is typically subjected to further decomposition, thus determining the emergence of a tree-like pattern for the wave function, as schematized in fig. \ref{fig1}.

The typical situation in which PSD is assumed to take place is the measurement process: At the end of a measurement the wave function of the universe (microscopic system + apparatus + rest of the universe) decomposes into $n$ elements, corresponding to the $n$ possible outcomes of the measurement. The elements are {\it non-overlapping} because they represent the apparatus with macroscopically different positions of the pointer, and they are {\it permanently} non-overlapping because IWE makes their spatial separation irreversible and permanent. By keeping in mind the measurement process, we can roughly identify three phases in the dynamics of PSD: (1) microscopic decomposition, (2) amplification, and (3) IWE.

(1) The decomposition of the wave function into channels originates on the microscopic level, mainly in the scattering processes, which have a structural tendency to generate such a form of decomposition. For example: (i) under suitable initial conditions a wave packet scattered by a potential barrier splits into a transmitted and a reflected wave packet. (ii) The wave function of an incident particle scattered by a massive particle splits into the scattered and the non-scattered part. (iii) Since states with a different number or type of particles can be naturally considered as non-overlapping, different scattering channels correspond to different channels in the sense of PSD. For example, a photon $\gamma$ scattered by an atom $a$ can ionize the atom or not:
\begin{equation}
|\gamma \ra |a \ra \to |\gamma' \ra |a' \ra + |e \ra |a^+\ra,
\end{equation}
where the state $|e \ra |a^+\ra$ represents the atom ionized and the electron emitted.

(2) In the amplification phase the microscopic system interacts with an amplifying system, typically an apparatus, and the spatial separation of the channels is amplified on a macroscopic level. Typical amplification processes occur in a photomultiplier or in a bubble chamber.

(3) In the third phase the amplifying system (the apparatus) interacts with the environment and the spatial separation of the channels transfers to the environment and becomes permanent, even though the channels of the apparatus or of the microscopic system successively overlap again. Here IWE has a slightly different feature than in decoherence theory; this can be explained in the case of the measurement process by utilizing the same notation as in the previous section. Assume as usual that immediately after the measurement, i.e., immediately after the amplification phase, the state of $\SS + \EE$ is $|S \ra |E \ra= (|S_1 \ra + |S_2 \ra)|E \ra$, where $|S_1\ra$ and $|S_2\ra$ are two non-overlapping wave packets corresponding to two macroscopically different configurations of the pointer of the apparatus. IWE is described by the following transaction:
\begin{equation} \label{pom2}
|S \ra |E \ra \to U(t) |S \ra |E \ra \approx |S_1 \ra |E_1(t) \ra + |S_2 \ra |E_2(t) \ra.
\end{equation}
Recall that equation (\ref{pom2}) holds true because wave packets are robust states in the collisional decoherence model, which is the most appropriate model in this case. In the usual decoherence scheme the states $|E_1(t) \ra$ and $|E_2(t) \ra$ are required to be {\it orthogonal } for $t$ greater than the decoherence time scale, while in the process of PSD they are required to be {\it non-overlapping}. This more stringent requirement is reasonable due to the presence of the amplification phase, which causes the states $|S_1 \ra$ and $|S_2 \ra$ to be different on the macroscopic level.

An experimental situation which well exemplifies the three phases of PSD is the beam splitter experiment: A particle is emitted by a source toward a beam splitter at an angle of $45^\circ$ with respect to the direction of the incident particle. After the beam splitter there are two detectors, one for the transmitted and one for the reflected particles. Microscopic decomposition occurs when the incident wave packet crosses the beam splitter, amplification occurs when the transmitted and the reflected wave packet are detected by the corresponding detector, and IWE occurs after detection.

\vspace{3mm}
PSD was discovered and studied by Bohm in connection with his ontological formulation of quantum mechanics \cite{Bohm:1951, Bohm:1951xx, Bohm:1987np,undivuni}. The reason why PSD is important in this formulation is the following. According to such a formulation, the particles of the universe follow a trajectory in configuration space defined by the guidance equation. After a measurement the actual trajectory enters and stays in one of the channels in which the wave function decomposes\footnote{Even though it has not been rigorously proved, the fact that the overwhelming majority of the Bohmian trajectories remain inside the non-overlapping channels of the wave function is very reasonable and is commonly assumed in the literature.}. Since the channels are permanently non-overlapping, the ``empty'' channels do not influence the actual trajectory, which evolves as though the other channels were collapsed. This is the so-called {\it effective collapse} of the wave function, which guarantees that the predictions of the de Broglie-Bohm theory correspond to those of standard quantum measurement theory, in which the ``empty'' channels are assumed to actually collapse.

Decoherence and PSD are not always clearly differentiated in the literature. For example, in \cite{Brown2005} one can read: ``This criticism is only sharpened by the recognition that decoherence --central to the modern view of Everett-- is also essential in the Bohmian picture. Although the Bohmian corpuscle picks out by fiat a preferred basis (position), the de Broglie-Bohm theory still has to tell some story about the measurement-induced effective collapse of the wavefunction. Bohm recognised this in 1952; from a modern perspective, we recognize this as the requirement that (1) decoherence occurs, and (2) the preferred basis which it picks out is (approximately) the position basis''. Another example: in Section 9.2 of the book \cite{bmbook}, the authors describe the process of effective collapse, and after this they claim: ``We have exactly the same effect in quantum mechanics, where it is called {\it decoherence}''. On the contrary, in \cite{baccia} the difference between PSD and decoherence is recognized: ``De Broglie-Bohm theory and decoherence contemplate two a priori {\it distinct} mechanisms connected to apparent collapse: respectively, separation of components in configuration space and suppression of interference. While the former obviously implies the latter, it is equally obvious that decoherence need not imply separation in configuration space''. In the previously introduced notation, this claim corresponds to the fact that if $|E_1(t) \ra$ and $|E_2(t) \ra$ are non-overlapping they are also orthogonal, but the opposite implication is not true.

It may be useful to discuss a concrete example in which decoherence, but not PSD, takes place. Consider the double-slit experiment and assume that immediately after crossing the two slits the particle interacts with an environmental photon. This interaction is represented by the transaction
\begin{equation}
(|\phi_1 \ra + |\phi_2 \ra ) |\gamma \ra \to |\phi_1 \ra |\gamma_1 \ra + |\phi_2 \ra |\gamma_2 \ra, 
\end{equation}
where $|\phi_i \ra$ is the part of the wave function of the particle which have crossed the slit $i=1,2$, and $|\gamma_i \ra$ represents the photon scattered by $|\phi_i \ra$. The states $|\gamma_1 \ra$ and $|\gamma_2 \ra$ are non-overlapping, but it is reasonable to admit that they spread and overlap under time evolution. Since also the two states $|\phi_1 \ra$ and $|\phi_2 \ra$ overlap again in the proximity of the photographic plate, the states $|\phi_1 \ra |\gamma_1 \ra$ and $|\phi_2 \ra |\gamma_2 \ra$ overlap under time evolution, and therefore PSD does not take place. However $|\gamma_1 \ra$ and $|\gamma_2 \ra$ remain orthogonal, thus decoherence takes place and the interference fringes do not appear on the photographic plate. This reflects the fact that, even though the environmental photon has recorded the slit crossed by the particle, there is no macroscopic observer aware of this information. In order to make this information accessible to a macroscopic observer it is necessary to amplify the difference between the two states $|\gamma_1 \ra$ and $|\gamma_2 \ra$ on a macroscopic level, and this leads unavoidably to PSD. With reference to the three phases of PSD, in this example we have microscopic decomposition and IWE, but not amplification. From this example it also follows that PSD rather than simple decoherence appears to be a more appropriate process for characterizing measurement-like interactions.

Let us therefore summarize the two main differences between decoherence and PSD: (a) decoherence theory only considers IWE, while PSD includes other two dynamical processes, namely microscopic decomposition and amplification; (b) in decoherence theory the states of the environment relative to orthogonal pointer states (e.g., the states $|E_1(t) \ra$ and $|E_2(t) \ra$ of equation (\ref{pom2})) are required to be {\it orthogonal}, while in PSD they are required to be {\it non-overlapping}.

\vspace{3mm}
If PSD is proven to be a real phenomenon, it certainly provides a simple and clear solution to the preferred-decomposition problem in the measurement process and in the MWI. By looking at fig. \ref{fig1} we see that the decomposition of the wave function into channels is evident at all times. There is a certain vagueness in regard to {\it where} the boundary between different channels has to be positioned, and also in regard to {\it when} a given channel can be considered as decomposed into sub-channels. This depends on the fact that PSD determines the emergence of a {\it pattern} of the wave function of the universe, and typically patterns are vaguely defined. Nevertheless they can be evident and real. For example, if the density $\eta(\x)$ of water vapor in the sky has a suitable pattern we see evident and well defined clouds, even though the boundary of a cloud cannot be exactly determined on a microscopic level. PSD would have to guarantee that the wave function $\Psi(x, t)$ of the universe has a pattern which allows us to ``see'' channels which are evident and well defined on a macroscopic level. See \cite{Wallace2003} for a conceptual discussion about the emergence of the worlds in the MWI as a pattern of the wave function of the universe.

One final remark is necessary: Unlike the pointer states of decoherence theory, the channels of the wave function of the universe can be determined independently of any decomposition of the universe into system and environment. This implies that PSD does not suffer, as decoherence does, from the problem of how to define such a decomposition in the context of the MWI.

\section{Formal representation of the pattern of permanent spatial decomposition} \label{formal} 

As mentioned previously, PSD determines the emergence of a pattern in the wave function of the universe, which is schematically shown in fig. \ref{fig1}. This pattern is vague, as patterns usually are, but nevertheless a formal representation of it can be given. Due to vagueness, various different kinds of representations arguably exist. The representation which I present here appears to me to be the most simple and natural. This section is a brief summary of parts of a previous paper of the author \cite{Galvan2009}.

Recall that a {\it decomposition} $\DD$ of a vector $\Psi$ has been defined as a finite or countable set of linearly independent vectors $\{\Psi_1, \Psi_2, \ldots \}$ such that $\sum \DD := \sum_i \Psi_i=\Psi$. The reason why the elements of a decomposition are required to be linearly independent will be explained shortly. The pattern shown in fig. \ref{fig1} is therefore represented by a suitable time-dependent decomposition $\{\DD_t\}_{t \geq 0}$ such that
\begin{equation} \label{scond1}
\sum \DD_t = U(t)\Psi_0,
\end{equation}
where $\Psi_0$ is the state of the universe at the initial time $t=0$. For simplicity the decompositions $\{\DD_t\}_{t \geq 0}$ will be assumed to be finite. The decompositions $\{\DD_t\}_{t \geq 0}$ have the following two properties: (a) their elements are permanently non-overlapping, and (b) for $t_2 \geq t_1$ the decomposition $\DD_{t_2}$ is a ``sub-decomposition'' of the decomposition obtained by letting $\DD_{t_1}$ evolve to the time $t_2$. These two properties will be now expressed in a formal way.

\vspace{3mm}
{\bf Property (a)}. A decomposition whose elements are permanently non-overlapping will be referred to as a {\it permanent spatial decomposition}. In order to formally define such a notion, the notions of {\it exact spatial decomposition} and of {\it (approximate) spatial decomposition} will first be defined. 

Let $\X$ denote the configuration space of the system, for example $\R^{3N}$; on $\X$ the projection valued measure $E$ is defined. Let $\XX:=\{\Delta_1, \ldots, \Delta_n\}$ be a finite partition of $\X$. The symbol $E(\XX)\Psi$ will denote the decomposition $\{E(\Delta_1)\Psi, \ldots, E(\Delta_n)\Psi\}$. A decomposition $\DD$ with $\sum \DD= \Psi$ is said to be an {\it exact spatial decomposition} if there exists a partition $\XX$ such that $\DD=E(\XX)\Psi$. The elements of an exact spatial decomposition are obviously {\it exactly} non-overlapping.

An approximate spatial decomposition can be defined as a decomposition which is approximately identical with an exact spatial decomposition. In order to formally define such a notion, let us consider the following tentative definition of a function $w$ of a generic decomposition $\DD:=\{\Psi_1, \ldots, \Psi_2\}$:
\begin{equation} \label{tent}
w(\DD)\stackrel{\hbox{\tiny tentative}}{:=}\inf_{\XX} \max_{1 \leq i \leq n} \left \{\frac{||\Psi_i - E(\Delta_i)\Psi||}{||\Psi_i||} \right \},
\end{equation}
where $\Psi=\sum \DD$, and $\XX$ ranges over the partitions of $\X$ with $n$ elements. We could therefore say that a decomposition $\DD$ is approximately a spatial decomposition if $w(\DD) \approx 0$. For reasons that will soon become clearer, it is better to generalize the definition of $w$ as follows. If $I$ is a subset of $\{1, \ldots, n\}$, define $\Psi_I:=\sum_{i \in I}\Psi_i$ and $\Delta_I:=\cup_{i \in I} \Delta_i$. We therefore define:
\begin{equation} \label{gendef}
w(\DD):=\inf_{\XX} \max_I \left \{\frac{||\Psi_I - E(\Delta_I)\Psi||}{||\Psi_I||} \right \}.
\end{equation}
If $w(\DD) \approx 0$ the decomposition $\DD$ is said to be an {\it approximate spatial decomposition}, and its elements are said to be {\it approximately non-overlapping}. For the sake of brevity, the attribute ``approximate'' will be omitted below.

As an example, consider a decomposition composed of two vectors $\{\Psi_1, \Psi_2\}$. In \cite{Galvan2009} it is shown that 
\begin{equation}
w(\{\Psi_1, \Psi_2\})=\frac{\left(\int \min\{|\Psi_1(x)|^2, |\Psi_2(x)|^2\} dx\right)^{1/2}} {\min\{||\Psi_1||, ||\Psi_2||\}}.
\end{equation}

Let us now define the notion of permanent spatial decomposition. Given the decomposition $\DD:=\{\Psi_1, \ldots, \Psi_n\}$, let $U(t)\DD$ denote the decomposition $\{U(t)\Psi_1, \ldots, U(t)\Psi_n\}$. Let $w^+(\DD)$ be defined as follows:
\begin{equation}
w^+(\DD):= \sup_{t \geq 0} w[U(t)\DD].
\end{equation}
If $w^+(\DD) \approx 0$ the decomposition $\DD$ is said to be a {\it permanent spatial decomposition} (PSD). 

As an example, consider a partition $\YY := \{\Sigma_1, \ldots, \Sigma_n \}$ of $\X$. A necessary and sufficient condition for $E(\YY)\Psi$ is a PSD is that, for any $t \geq 0$, a partition $\XX_t := \{\Delta_1^t, \ldots, \Delta_n^t \}$ of $\X$ exists such that
\begin{equation}
\frac{||U(t) E(\Sigma_I) \Psi - E(\Delta_I^t)U(t)\Psi||}{|| E(\Sigma_I) \Psi||} \approx 0 \hbox{ for any } I \subseteq \{1, \ldots, n\}.
\end{equation}

Returning to property (a), it can be expressed by the following condition:
\begin{equation} \label{scond2}
w^+(\DD_t) \approx 0 \hbox{ for } t \geq 0.
\end{equation}

\vspace{3mm}
{\bf Property (b)}. In order to define this property it is necessary in some way to define that the decomposition $\DD_{t_2}$ is ``finer'' than the decomposition $U(t_2-t_1) \DD_{t_1}$ (note that both the decompositions are decompositions of the same vector $U(t_2)\Psi_0$). We therefore say that a decomposition $\DD'$ is {\it finer} than a decomposition $\DD$ if there exists a map $h:\DD' \to \DD$ such that 
\begin{equation}
\Psi_i=\sum_{\Psi'_j \in h^{-1}[\Psi_i]} \Psi'_j \hbox{ for any } \Psi_i \in \DD.
\end{equation}
If $\DD'$ is finer than $\DD$, we write $\DD' \preceq \DD$. In words, $\DD' \preceq \DD$ iff $\DD'$ and $\DD$ are decompositions of the same vector, and every element of $\DD$ is the sum of a different subset of elements of $\DD'$.

It is easy to prove that: (i) if the map $h$ exists it is univocally defined and surjective; (ii) $\preceq $ is a partial order; (iii) $\DD' \preceq \DD$ implies that $\sum \DD' = \sum \DD$ and $w(\DD') \geq w(\DD)$. Point (i) derives from the fact that the elements of a decomposition are linearly independent, and this is the reason why this requirement has been included in the definition of decomposition. Moreover the second implication of point (iii) holds true because the general definition (\ref{gendef}) has been chosen for $w(\DD)$ in place of (\ref{tent}).

Property (b) can therefore be expressed by the following condition:
\begin{equation} \label{scond3} 
\DD_{t_2} \preceq U(t_2 - t_1) \DD_{t_1} \hbox{ for } t_2 \geq t_1.
\end{equation}

\vspace{3mm}
According to the formal definition of properties (a) and (b) it is natural to define a {\it tree structure} for $\Psi_0$ as a time-dependent decomposition $\{\DD_t\}_{t \geq 0}$ satisfying the conditions (\ref{scond1}), (\ref{scond2}), and (\ref{scond3})\footnote{These conditions are actually partially overlapping. Non-overlapping conditions are the following: (1) $\sum \DD_0=\Psi_0$, (2) $\DD_{t_2} \preceq U(t_2 - t_1) \DD_{t_1} \hbox{ for } t_2 \geq t_1$, and (3) $w(\DD_t) \approx 0$ for $t\geq 0$.}. A suitable tree structure formally represents the pattern of the wave function of the universe determined by PSD; of course this tree structure is only vaguely defined by the pattern, in the sense that (infinitely) many tree structures can represent the same pattern.

\section{Discussion} \label{conclusion}

In this paper the emphasis has been placed on the process of {\it permanent spatial decomposition} (PSD), i.e., the (hypothesized) property of the wave function of the universe of decomposing continuously into permanently non-overlapping parts. This process has been analyzed and compared with decoherence, showing that: (i) PSD is basically different from decoherence, because in the latter process only the interaction with the environment is considered, while the former also consider two processes preceding the interaction with the environment, namely microscopic decomposition and amplification. (ii) PSD, but not decoherence, provides a complete solution to the problem of how to single out a preferred decomposition of a suitable state vector into the sum of a finite or countable set of vectors; this allows PSD to provide a simple and non-elusive solution to the problems of how to decompose the state of microscopic system + apparatus after a measurement and how to define the worlds in the MWI.

There are various interpretations or formulations of quantum mechanics which are strictly connected with PSD: These of course include the MWI, but also the de Broglie-Bohm formulation, for the reasons explained in section \ref{process}. Furthermore, two others have recently been proposed \cite{galvan:4, zanghimw}: The first one is a proposal by the author according to which the particles of the universe follow a trajectory constrained to remain inside a channel of the wave function; this constraint is expressed mathematically by means of a new principle, the {\it Quantum Cournot principle}, which can be shown to be a generalization of the Born rule. The second proposal is a version of the MWI in which a mass ontology is added to the wave function, and in which the worlds are explicitly defined as channels of the wave functions.

PSD is however independent of any interpretation/formulation, in the following sense: While the existence of an ontology such as trajectories or the mass ontology must be postulated and may be questioned, PSD has an objective character, i.e., at least in principle it can be deduced (or rejected) by only studying Schr\"odinger's evolution (and assuming non-cospirative initial conditions). The conceptual nature of PSD is therefore very different from the related interpretations/formulations.

As mentioned previously, PSD is mainly known and considered in the rather restricted community of Bohmian physicists, and it is probably ignored by the majority of the physicists interested in subjects related to the preferred-basis problem, such as the measurement problem or the MWI. Moreover, it appears to me that in the community of Bohmian physicists the emphasis is placed on the ontology (e.g., trajectories or mass ontology) rather than on PSD, which is often assumed without further discussion as a natural property of the wave function (I also assumed this in my previous papers \cite{galvan:1,galvan:4}). I now argue on the contrary that PSD is an important subject which deserves to be studied by itself, independently of any formulation/interpretation, and which deserves to be brought to the attention of the physicists considering decoherence to be the only solution to the preferred-basis problem.

\bibliographystyle{custom}
\bibliography{general}
\end{document}